\documentclass{elsart}

\usepackage{epsfig}

\usepackage{amssymb}

\begin{document}

\begin{frontmatter}



\title{Multifractal nature of stock exchange prices}
\author{M. Ausloos$^1$ and K. Ivanova$^{2}$} 
\address{$^1$SUPRAS and GRASP,
B5, University of Li$\grave e$ge, B-4000 Li$\grave e$ge,
Euroland \\  $^{2}$ Pennsylvania State University, University Park, 
PA 16802, USA}
\begin{abstract} The multifractal structure of the
temporal dependence of the Deutsche Aktienindex (DAX) is analyzed. 
The $q$-th order moments of the structure functions
and the singular measures are calculated.  The generalized Hurst 
exponent $H(q)$ and the $h(\gamma(q))$ curve indicate a hierarchy 
of power law exponents. This approach leads to characterizing the 
nonstationarity and intermittency pertinent
to such financial signals, indicating differences with turbulence data.  
A list of results on turbulence and financial markets is presented for
asserting the analogy. 
\end{abstract}
\end{frontmatter}
\section{Introduction}
The analogy between fluid turbulence and financial markets has been previously noticed [1,2]. 
The energy
flow in hydrodynamics is supposed to mimic the information transfer in financial
markets, or the entropy variation of the market \cite{nvmafex}. Some hierarchical
structures are thought to exist,
leading to cascades of information and clusters of buy-sell orders  \cite{AMS98a,5} 
and sometimes to crashes \cite{6}. Surprisingly, more work on ''financial cascades'' can be
found on the foreign exchange market time series [7-14] rather than on share prices 
or stock market indices. However, since all such financial time series indicate 
nonstationarity, a multifractal \cite{HP83,mf} description seems useful [13,14,17-24].
The goal of this paper is  to present original results on the Deutsche Aktienindex (DAX), 
obtaining  the roughness parameter ($H_1$) and the degree
of intermittency ($C_1$). The $q$-th
order moments of the structure functions and the singular measures are
constructed thereof [23-25].
The behaviors are consistent with the multi-affine properties of other
turbulent phenomena \cite{kiandta}. Understanding
the processes that underlie these "macroscopic" effects remains at the speculative level. 
Numerical differences are pointed out. Nevertheless, the nonexhaustive list of references 
is a sufficient argument for asserting the analogy and suggesting further work.
\section{Experimental data analysis}
The Deutsche Aktienindex (DAX) data used here is from $http://deutsche-boerse.com/$ . 
It goes from  Oct. 01, 1959 till
Dec. 30, 1998. Let the data consist in a series $y(t)$, where $t$ is a discrete variable $t_i$,
i.e. $N= 9818$ data points.  First, the scaling range (if any) is found from a detrended 
fluctuation analysis ($DFA$) method [27]. It is found that  the scaling range  extends up to 
256 days, after which
the error bars become too large. The scaling (Hausdorff) exponent $Ha \sim 0.54$  is about
the same as that of the DJIA \cite{kimalev}.  It is expected that the self-affine fractal 
dimension $D$ = $2 - Ha$.
\begin{figure}[ht] \begin{center} \leavevmode \epsfysize=7cm \epsffile{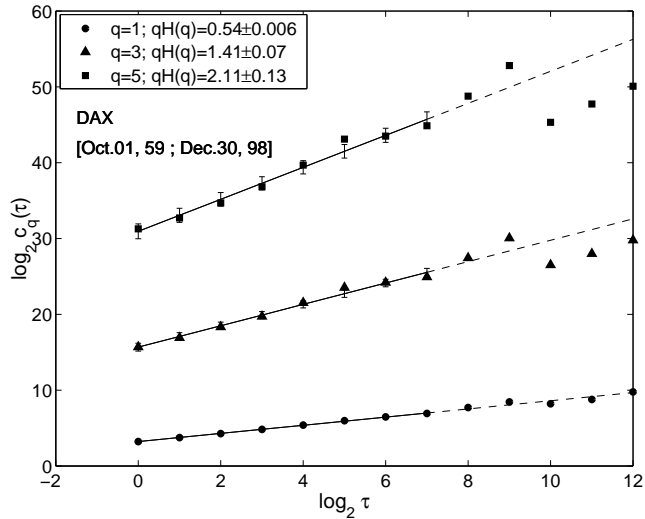} 
\caption{The DAX q-th order structure functions $c_q(\tau)$
as a function of time lag $\tau$ during Oct. 01, 59 - Dec. 30, 98} \end{center}
\end{figure}
Knowing the scaling range, one can reconsider the problem of the time variation of $D$ 
(or $Ha$). In the multifractal approach \cite{kimalev,davis,kiandta}, one seeks for 
various scales of self-affinity, through the so-called $q$-th order structure functions
\begin{equation} c_q(\tau) = \left<|y(t_{i+r}) - y(t_i)|^q \right>_{\tau} \qquad
i=1,2, \dots , N -r \end{equation} where only non-zero terms are considered in
the average ${\langle ... \rangle}_{\tau}$ taken over all $N-r$ couples $(t,t')$
such that $\tau = |t-t'|$ is a characteristic time lag, $\tau=t_{i+r}-t_i$, with 
$r \ge 0$, see Fig.1. Assuming a power law dependence of the structure function 
$c_q(\tau)$, the $H(q)$ spectrum is
defined through \begin{equation} c_q(\tau)\sim \tau^{qH(q)} \qquad  q\ge 0, \end{equation} 
and is shown in Fig.2, as $qH(q)$; note that $Ha=H(1)$ [22].
The {\it intermittency} of the signal can be studied through a singular measure analysis. 
Define a measure $\varepsilon(1;i)$ as 
\begin{equation} \varepsilon(1;i)=\frac{|\Delta y(1;i)|}{<\Delta y(1;i)>}, \qquad
i=0,1, \dots, N -1 \end{equation} where $\Delta y(1;i)=y(t_{i+1})-y(t_i)$ is the
small-scale gradient field and
\begin{equation}<\Delta y(1;i)>= \frac{1}{N}\sum_{i=0}^{N-1}|\Delta y(1;i)|. \label{ave} \end{equation}
\begin{figure}[ht] \begin{center} \leavevmode \epsfysize=7cm \epsffile{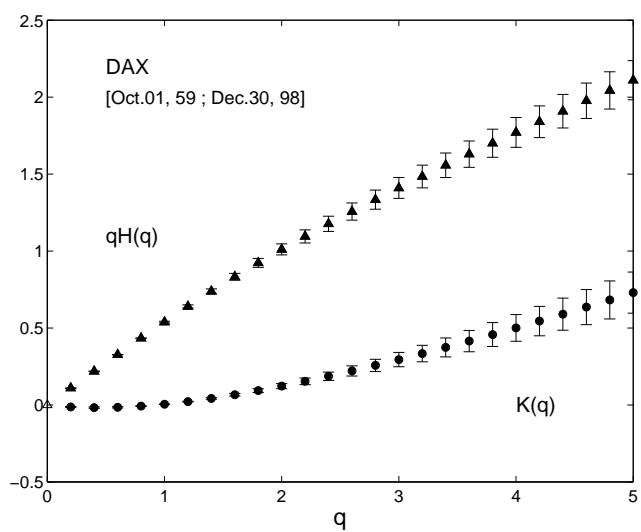} 
\caption{The DAX $qH(q)$ and $K(q)$ scaling functions for 
 [Oct. 01, 59-Dec. 30, 98]} \end{center} \end{figure} 
Next, define a series of ever more coarse-grained and ever shorter fields 
$\varepsilon(r;l)$ where $0<l<N-r$ and $r=1,2,4,\dots,N=2^m$. The average measure
in the interval $[l;l+r]$ is
\begin{equation}\varepsilon(r;l)=\frac{1}{r}\sum_{l'=l}^{l+r-1} \varepsilon(1;l')
\qquad l=0, \dots , N - r \end{equation} The scaling properties are then searched for through 
$\chi_q(\tau)=<\varepsilon(r;l)^q>_{\tau}\sim\tau^{-K(q)}$ for $q\ge 0$. 
Thereby, the multifractal
properties of the DAX signal are expressed by two scaling functions, $H(q)$ for
describing the roughness of the signal and $K(q)$ its intermittency. 
The $K(q)$ spectrum (Fig.2) is closely related to the generalized dimensions 
$D_q=1-K(q)/(q-1)$ [15,16]. A nonlinearity of both $qH(q)$ and $K(q)$  implies 
multifractality. Let $C_1 = - \left| {dK(q)/dq} \right|_{q=1}$.   It seems to be 
a measure of the information entropy of the system \cite{nvmafex}. For the DAX, 
$C_1 =  0.07\pm0.002$, 
interestingly compared to $C_1=0.27$ for the DJIA \cite{kimalev}. Note that both 
$Ha$'s = $0.54\pm0.006$ are similar.
\begin{figure}[ht] \begin{center} \leavevmode \epsfysize=7cm \epsffile{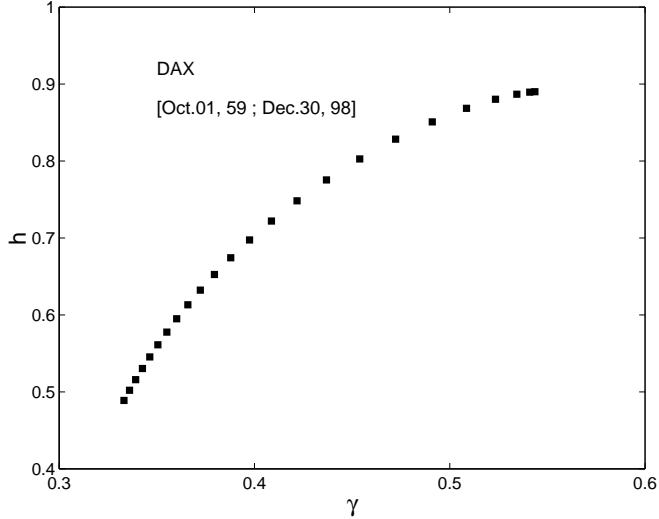} 
\caption{The $h(\gamma)$ curve for the DAX data from Oct.
01, 1959 to Dec. 30 1998} \end{center} \end{figure} 
Let $ \gamma(q)= {d(qH(q))}/{dq} $ and $ h(\gamma(q))=1+q\gamma(q)-qH(q)$.  
The function $h(\gamma(q)) $ is the fractal
dimension of the Cantor dust set of points having the same roughness exponent 
$\gamma(q)$ [16]. The $h(\gamma)$ function acts like the function $f(\alpha)$ 
in Ref.[16]. The $h(\gamma(q))$ curve is shown in Fig. 3. The error bars
are easily estimated from Fig.1 and the above equations.  
Note that $h(\gamma(q))$ reaches a 
maximal dimension for some finite $\gamma_0$ corresponding to the 
fractal dimension of the signal if it was self-affine [22].

\section{Discussion}
For a long time, the only available theoretical background to the statistical 
behavior of prices was the Efficient Market Theory (EMT).  The EMT has been 
usually identified with the random walk character of prices ($Ha=0.5$, $C_1 =0$).  
In finance papers, $Ha > 0.5$ values have been reported , -the deviations from
$0.5$ being explained by regulations imposed on the market by central authorities.  
However, $Ha$ seems to be often different from $0.5$, and $C_1$ is finite. Thus, one 
should expect multifractal features indicating that the origin of scaling laws
in financial time series is to be found in exogenous forces that cover a wide variety of influences.  
Finally, let us conclude that fluctuations in DAX and other financial market
data time series \cite{h1c1,kimalev} can be compared to those occurring in turbulence. 
Analogies with
intermittency, cascades, period doubling \cite{18} can be invoked with
Kolmogorov $1/3$-law process \cite{K62} and the fractional Brownian motion as 
basic ideas. Recall  that intermittency as introduced in turbulence led to a 
multifractal description, with $Ha=1/3$ and $C_1 \sim 0.05$.The above data 
indicate that the DAX (and DJIA [9]) $H_1 \sim 0.5$ and $C_1 \sim 0.07$ are
far away from the turbulence domain $H_1 \sim 0.3$ and $C_1 \sim 0.05$. 
Thus there is some analogy, but models should be different!
\vskip 0.2cm
{\bf Acknowledgements} \vskip 0.2cm Scientific computing is greatly indebted to D.
Stauffer. So are this work and the authors. 

{\it Note added in proofs :}

A surrogate data analysis has been performed as well. The corresponding values of $qH(q)$
as found in Fig.1 are $0.05\pm0.005$; $0.12\pm0.03$; $0.19\pm0.06$, respectively.

\end{document}